\documentclass[prl, twocolumn, preprintnumbers, floatfix]{revtex4}

\usepackage{graphicx}

\begin{document}

\preprint{LA-UR-13-26796}

\title{Creation of matter wave Bessel beams}

\author{C. Ryu}
\author{K. C. Henderson}
\author{M. G. Boshier}
\affiliation{Physics Division,Los Alamos National Laboratory, Los Alamos, NM 87545, USA}

\begin{abstract}
Bessel beams are plane waves with amplitude profiles described by Bessel functions.  They are important because of their property of limited diffraction and their capacity to carry orbital angular momentum.  Here we report the creation of a Bessel beam of de Broglie matter waves.  The Bessel beam is produced by the free evolution of a thin toroidal atomic Bose-Einstein condensate (BEC) which has been set into rotational motion.  By attempting to stir it at different rotation rates, we show that the toroidal BEC can only be made to rotate at discrete, equally-spaced frequencies, demonstrating that circulation is quantized in atomic BECs.  The method used here to generate matter wave Bessel beams with a “Painted Potential” can be viewed as a form of wavefunction engineering which might be extended to implement arbitrary cold atom matter wave holography.
\end{abstract}

\maketitle

Bessel beams - plane waves with amplitude profiles described by Bessel functions – are important because of their property of limited diffraction \cite{Durnin1987} and their capacity to carry orbital angular momentum \cite{Volke-Sepulveda2002}.  These characteristics have already been exploited to improve the performance of interferometry \cite{Fortin2004}, microlithography \cite{Erdelyi1997}, and imaging \cite{Ok2013}.  Bessel beams have also been used for micromanipulation \cite{Arlt2001}, where it has been proposed that they might create “tractor beams” \cite{Chen2011}.  To date, Bessel beams have been realized with light \cite{Durnin1987, McGloin2005} and with sound \cite{Lu1990}, and related Gauss-Laguerre beams with orbital angular momentum have also been realized with electrons \cite{Uchida2010, Verbeeck2010, McMorran2011}.  Here we report the first creation of a Bessel beam of de Broglie matter waves.  The Bessel beam is produced by the free evolution of a thin toroidal atomic Bose-Einstein condensate (BEC) which has been set into rotational motion.  By attempting to stir it at different rotation rates, we show that the toroidal BEC can only be made to rotate at discrete, equally-spaced frequencies.  This is a direct demonstration that circulation is quantized in atomic BECs.  It also dictates that atoms in the Bessel beam must carry quantized orbital angular momentum, and it means that the preparation of beams with this method does not require precise control of the initial BEC rotation rate.  We show that the results of the experiment are consistent with a simple theory.  The method used here to generate matter wave Bessel beams with a “Painted Potential” \cite{Henderson2009} can be viewed as a form of wavefunction engineering which might be extended to implement arbitrary cold atom matter wave holography.

The useful properties of Bessel beams follow from the result that if ${J_n}(\rho )$  is the $n$th Bessel function of the first kind, then  ${\nabla ^2}{J_n}(\rho ){e^{ - i n \phi }} =  - {J_n}(\rho ){e^{ - i n \phi }}$.  Consequently ${J_n}(\rho ){e^{ - i n \phi }}$  satisfies the Helmholtz equation and so a plane wave with this amplitude profile and phase winding propagates without any change in form.  While this perfectly non-diffracting plane wave has infinite transverse extent, practical implementations of these waves are still useful because the effect of finite extent is to limit the non-diffracting region to finite length \cite{Durnin1987}.  In the particular case of quantum mechanics, $\psi (\rho,\phi,t) = {J_n}(\alpha  \rho ){e^{ - i n \phi }}{e^{ - i\hbar {\alpha ^2}t/2m}}$  (with $\alpha$ being a constant) is a diffraction-free solution of the free Schr{\"o}dinger equation for a particle of mass $m$.  To make experimentally an approximation to this wavefunction, we note that at long times the solution of the 2D free Schr{\"o}dinger equation has a \lq\lq{}Fraunhofer diffraction limit\rq\rq{} solution
\begin{equation}
\label{eq:Fraunhofer}
\psi ({\bf{r}},t) = A{e^{i\frac{{m{r^2}}}{{2\hbar t}}}}\frac{{\Phi (m{\bf{r}}/\hbar t,t = 0)}}{{\hbar t/m}}
\end{equation}
where $A$ is a constant, $\Phi ({\bf{k}},t = 0)$  is the 2D momentum space wavefunction at time $t$, and the position vector ${\bf{r}} = (x,y)$.  Since a spatial wavefunction in the form of a thin torus of radius $R$  with phase winding number $n$, $\psi (r,\phi ) = \delta ( r - R) {e^{ - i n \phi }}$, has $\Phi (k,0) = {J_n}(kR)$, free expansion of such a wavefunction will create an approximation to the desired Bessel wave.

\begin{figure}
\includegraphics[trim=0.5in 0 0 0, width=3.50in]{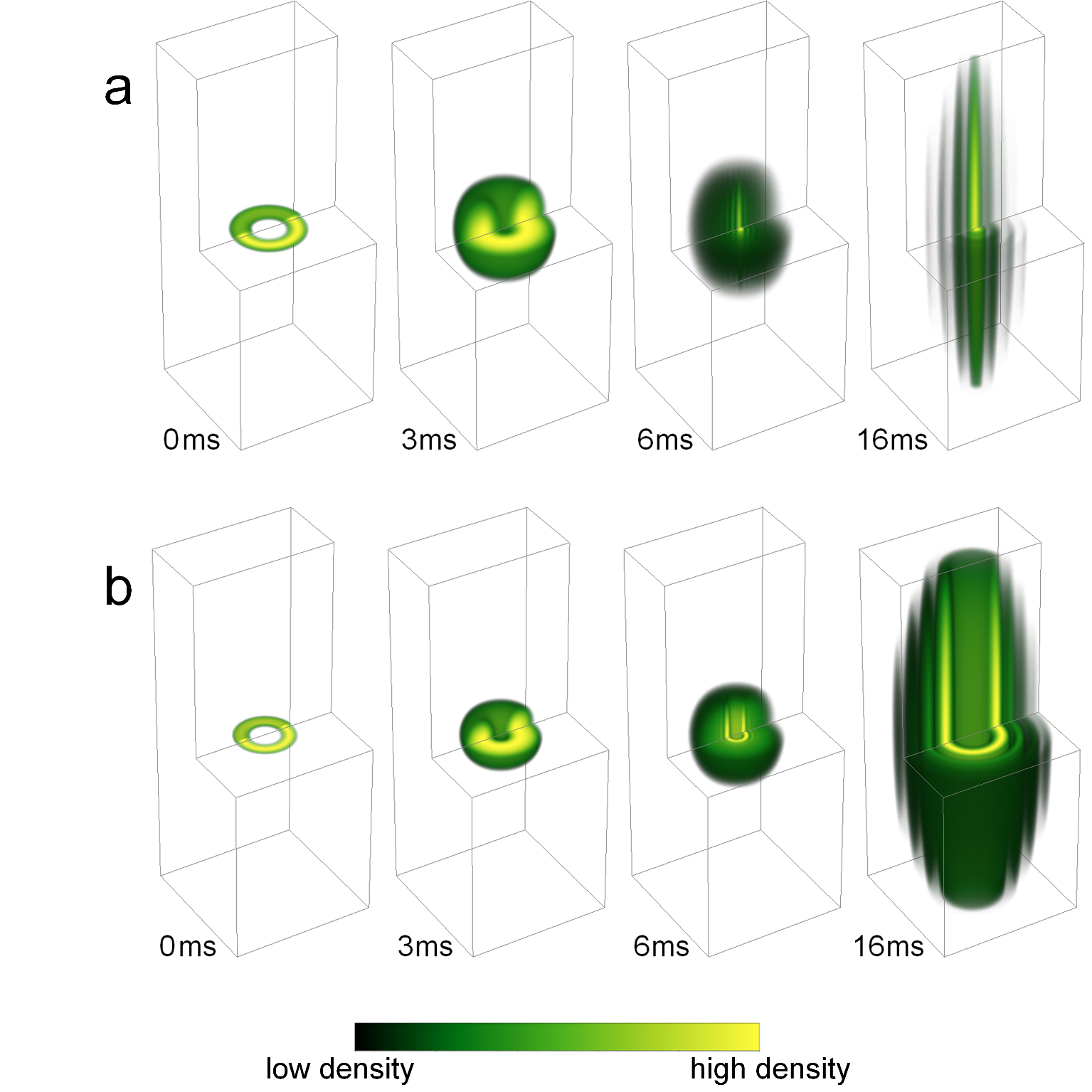}
\caption{\label{fig:GPE-3D} Numerical simulation of toroidal BEC evolving into a Bessel Beam.  Volume-rendered density distributions obtained at the times shown from simulations of free expansion of the toroidal Bose-Einstein condensate, showing how a Bessel beam is formed.  (a) winding number $n=0$ and (b) $n=5$.  Solid boundary lines show how one quarter of the rendered volume has been clipped to reveal the interior structure.  Dimensions of the rendered volume are 50\,$\mu$m$\times 50\,\mu$m$\times 100\,\mu$m.}
\end{figure}

Our experiment follows this recipe with a toroidal atomic Bose-Einstein condensate (Fig.\,\ref{fig:GPE-3D}).  In this system the requirement for a single integer phase winding number is guaranteed by the quantized rotational response that follows from the superfluid being described by a complex-valued order parameter $\psi$.  The associated superfluid velocity is $\textbf{v} = \frac{\hbar}{m} \nabla \phi$, where $\phi$  is the phase of $\psi$ and $m$ is the mass of the particles forming the superfluid current.  The line integral known as the circulation then takes on quantized values,
\begin{equation}
\label{eq:circulation}
\oint \textbf{v} \cdot d\textbf{l}  = \frac{\hbar }{m}\Delta \varphi  = n\frac{h}{m}
\end{equation}
because the requirement that the order parameter be single-valued restricts the phase change  around the path to $2 \pi n$, where $n$ is the (integer) winding number.  For a thin toroid of radius $R$ these allowed states have quantized angular velocity $\Omega  = n {\Omega _0}$, where ${\Omega _0} =  \hbar /m{R^2}$, and quantized orbital angular momentum $n \hbar$ per atom.  In addition, these states are also metastable persistent current states with lifetimes of several seconds \cite{Ryu2007}.

\begin{figure}
\includegraphics[width=3.50in]{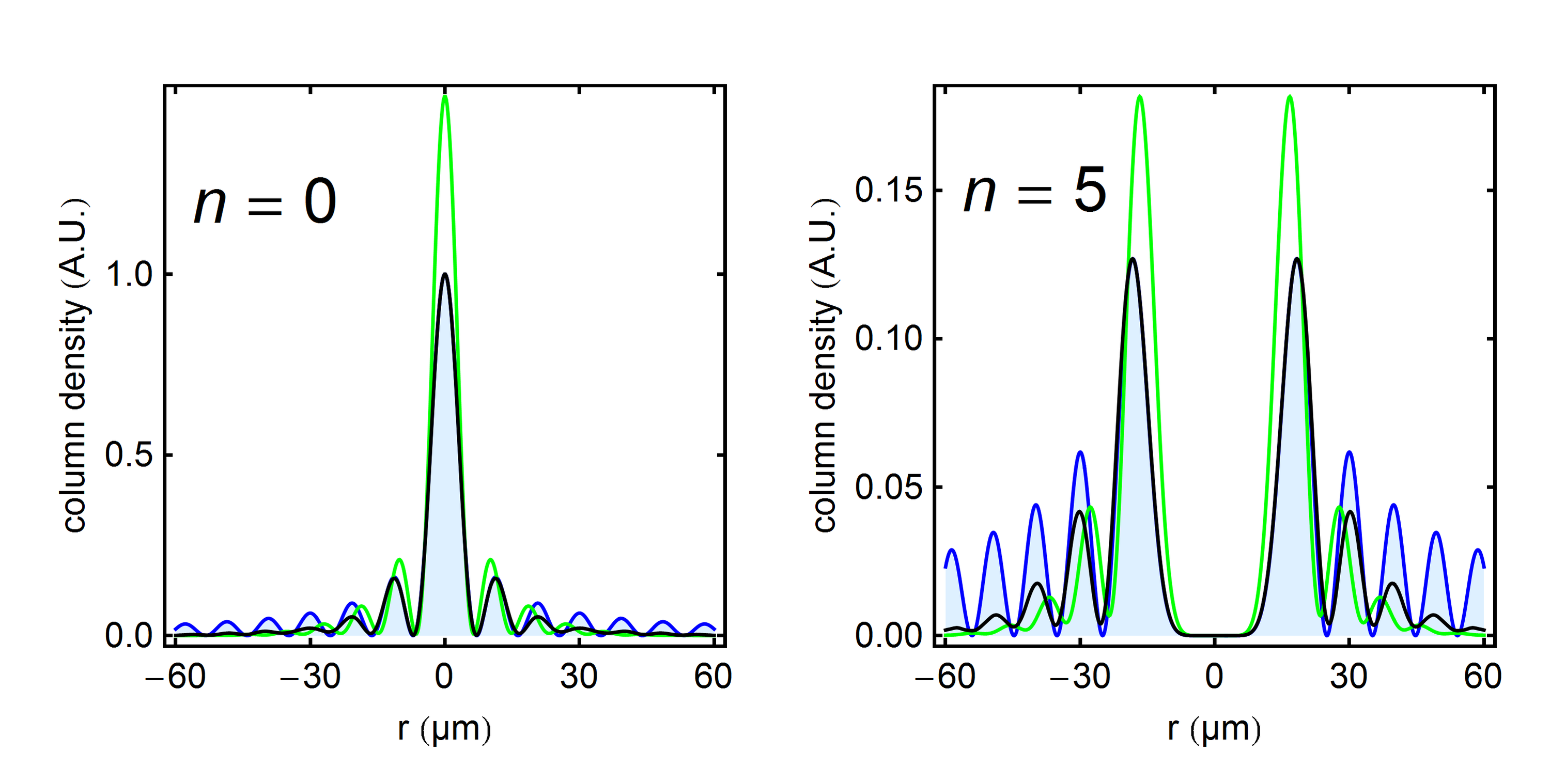}
\caption{\label{fig:GPE-radial} Effect of interactions and finite width of torus on Bessel approximation.  Black curve shows (for $n=0$ and $n=5$) the column density obtained at 34\,ms expansion time from a full 3D GPE simulation of the expansion dynamics of toroidal BECs with the parameters appropriate to the experiment.  Blue filled curve shows the squared Bessel function $J_{n}(r)^{2}$ overlaid after $\sim10\%$ expansion in radial scale due to effect of interactions.  Green curve shows the GPE simulation for a BEC without interactions plotted using the same radial scaling.}
\end{figure}

The discussion so far has assumed a thin toroidal BEC obeying the linear Schr{\"o}dinger equation. However a real system is 3D, and the strong interatomic interactions lead to dynamics described by the nonlinear Gross-Pitaevskii equation (GPE).  Fortunately we find from numerical solutions of the GPE that these factors should not cause a serious deviation from ideal Bessel evolution.  Fig.\,\ref{fig:GPE-3D}, computed for the parameters of our experiment, shows the free expansion of toroidal BECs with $n=0$ and $n=5$.  The toroidal condensate initially expands in all directions ($t=3\,$ms) until the wavefunction reaches the center of the toroid ($t=6\,$ms) where the interference is either constructive ($n=0$), forming a central peak, or destructive ($n\ge1$), forming a central hole.  The BEC then evolves into a long thin condensate with concentric layers structured in accordance with the relevant Bessel function ($t\ge16\,$ms).  Fig.\,\ref{fig:GPE-radial} shows that the column density along the axial direction has fringes that match up well with the corresponding Bessel function, the main difference being that the fringe amplitude falls off with radius faster for the GPE solution.  It is clear from the difference between the black and green curves that interactions only slightly expand the fringe pattern.  The main effect of interactions is seen in the axial direction, where they drive the rapid expansion that leads to the elongated beam-like object depicted in Fig.\,\ref{fig:GPE-3D}.

Our experimental procedure to create rotating thin toroidal BECs is conceptually straightforward:  we make an $^{87}$Rb condensate in the $5^2S_{1/2}$ $|F = 2, m_F = 2\rangle$ ground state by evaporating into a rotating toroidal trap, and we then measure the resulting superfluid velocity by releasing the condensate and imaging it after a period of free expansion.  The rotating toroidal traps are realized using the "Painted Potential" technique \cite{Henderson2009} for creating condensates in arbitrary time-averaged optical dipole potentials formed by painting with a rapidly moving tweezer beam on top of a tight horizontal sheet trap.  The technique can realize and animate any 2D potential that can be drawn on a sheet of paper.  The trap diameter is 19$\,\mu$m, for which ${\Omega _0} \approx 2\pi  \times 1.3\,$rad/s, and the full width at half-maximums (FWHMs) of the BEC density in the axial and radial directions are ~2$\,\mu$m and ~5$\,\mu$m respectively for the typical atom number of 40,000.   

For the experiments reported here, the system of Ref. \cite{Henderson2009} has been improved in three ways.  First, the horizontal light sheet is now more uniform because it is created by scanning a spherically focused beam instead of the static cylindrically-focused beam used previously.  Second, the intensity of the tweezer beam which paints the potential is now modulated synchronously with the painting scan to improve the flatness of the potential.  Third, separate lenses are used for focusing the tweezer beam and for imaging the trapped condensate, delivering better performance in both areas.  The tweezer beam (waist =$ 11\,\mu$m and wavelength $\lambda = 1064\,$nm) paints at $10\,$kHz scan frequency a $19\,\mu$m diameter torus with trap depth $240\,$nK and radial trapping frequency $120\,$ Hz.  The tweezer beam intensity is modulated so as to create a rotating barrier of height 45\,nK and FWHM thickness $13\,\mu$m (the scheme was inspired by the proposal of \cite{Brand2001}, although our torus is too thin to easily support defects inside the ring).  The horizontal sheet optical dipole trap providing the vertical confinement has trapping frequency 400\,Hz.  The experimental timing is as follows.  The optical dipole potential, superimposed on the cold atoms in the magnetic trap, is ramped up over 100\,ms, at which point the magnetic trap is turned off.  After a 200\,ms equilibration time the trapped atoms are evaporated to condensation by lowering the sheet intensity over 3\,seconds, forming a condensate which is rotating with the trap.  A repulsive barrier with height greater than the chemical potential is initially painted across the toroid to force the cold gas to rotate with the trap as it condenses.  After the BEC has formed in the rotating trap, the barrier is lowered adiabatically over 200\,ms, transforming the potential into toroidal form.  After 100\,ms of equilibration the trapping potential is turned off and the condensate is imaged in absorption along its long axis following 34\,ms of free expansion.

\begin{figure*}
\includegraphics[width=7.0in]{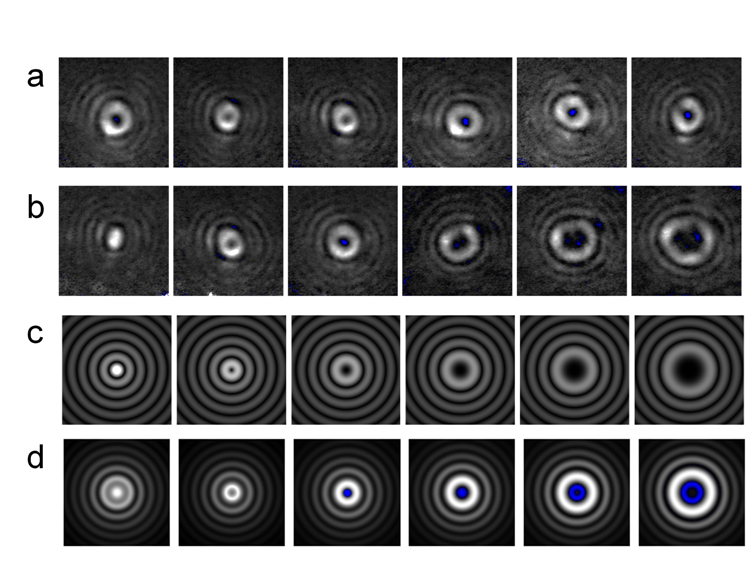}
\caption{\label{fig:Data} Experimental Bessel beam images and theoretical models.  (a) and (b) show time of flight absorption images taken after 34\,ms of free expansion.  The dimensions of each image are 100\,$\mu$m$\times 100\,\mu$m.  White (black) corresponds to regions of largest (smallest) absorption.  Blue indicates negative absorption, which is small and a result of imperfect background subtraction, except in the center of the images where a large apparent negative absorption is due to diffraction of the probe beam (see discussion in text).
(a) Six consecutive runs under nominally identical conditions, (b) Images showing winding numbers from $n=0$ to $n=5$, (c) Squared Bessel functions $J_{n}(r)^{2}$ for $n=0$ to $n=5$, (d) Expected images for $n=0$ to $n=5$ computed using the imaging model discussed in the text.}
\end{figure*}

\begin{figure}
\includegraphics[width=3.0in]{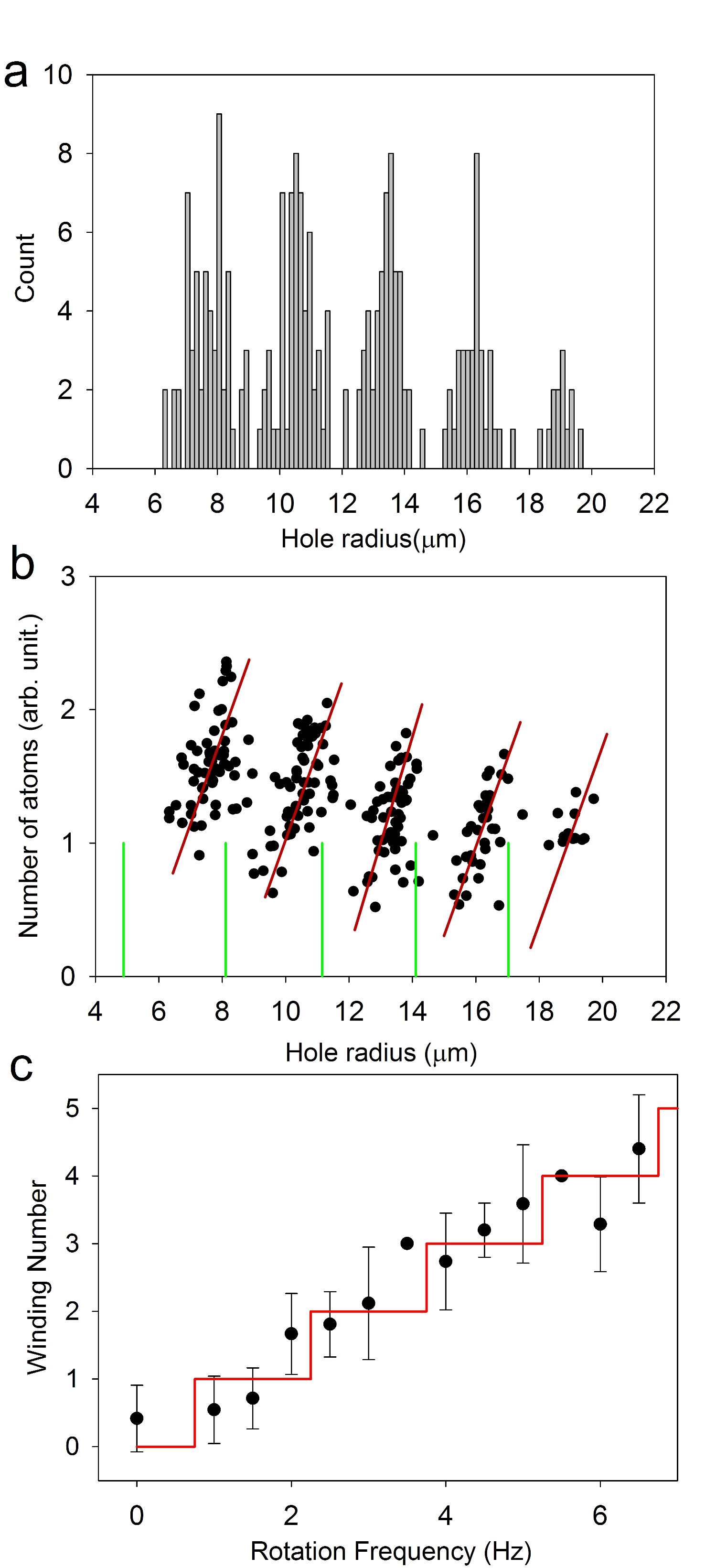}
\caption{\label{fig:Analysis} (a) Histogram of fitted values for radius of first interference fringe when the BEC is stirred over closely-spaced frequencies up to 7\,Hz.  The five groups correspond to winding number $n=1$ to $n=5$.
(b) Fringe height versus fringe radius scatter plot of the same data set.  Green lines show the fringe radii expected from the simple model of non-interacting atoms in an infinitely thin torus.  Red lines are drawn to guide the eye.  (c) Measured winding number versus rotation frequency of the trap in which the toroidal condensate was created. The error bars represent the standard deviation of the distribution of winding numbers measured at each trap rotation frequency.  The red line shows the theoretical prediction for an ideal thin toroidal condensate.}
\end{figure}

Fig.\,\ref{fig:Data}(a) and (b) show time of flight (TOF) images of the column density obtained in this way, exhibiting concentric ring patterns that agree qualitatively with the Bessel function intensity distributions ${\left| {{J_n}\left( {\frac{{mR}}{{\hbar t}}r} \right)} \right|^2}$ predicted by Eq.\,(\ref{eq:Fraunhofer}) and shown in Fig.\,\ref{fig:Data}(c).  Imaging these beam-like optically thick condensates is non-trivial because diffraction and refraction alter the intensity profile of the probe beam as it propagates through the condensate.  We have developed a model of this process which predicts the simulated absorption images shown in Fig.\,\ref{fig:Data}(d) for our conditions.   Modeling of the experimental images must take into account absorption, lensing, and diffraction of the probe beam as it propagates through the condensate.  Our current model for this process proceeds as follows:  First, the simulation of the free expansion dynamics is done in two stages: the GPE is solved numerically in polar coordinates until a time where the interaction energy of the condensate has become negligible, after which the wavefunction is projected onto a Cartesian mesh and propagated directly to any desired time using a decomposition into plane waves.  Second, an appropriate two-level atom susceptibility is applied to the corresponding density distribution to model the medium through which the probe beam will propagate.  A plane wave propagated through that medium using the paraxial wave equation yields the electric field distribution of the imaging beam at the exit of the condensate.  That field is then propagated numerically through free space to the focal plane of the imaging system to obtain a simulation of the image seen by the camera.  Full details of this procedure and the complete results will be published elsewhere, but the important result is that there exists a focus plane for the imaging system where absorption images appear very similar to the condensate column density.  One notable deviation from the exact column density is that diffraction following absorption on entry to the condensate causes the intensity of the probe beam to actually increase above its input value \cite{Boshier1982}, explaining the negative absorption seen at the center of many of the images in Figs.\,\ref{fig:Data}(a) and (b) with $n\ge2$.  The agreement between these experimental images and the corresponding predictions of the imaging model [Fig.\,\ref{fig:Data}(d)] is qualitatively good.  The imaging model used here makes the paraxial approximation and it also assumes perfect imaging optics.  We expect that a more sophisticated imaging model now under development will yield quantitative agreement between model and data, and also explain the apparent correlation between fringe radius and atom number seen in Fig.\,\ref{fig:Analysis}(b).

The simple classical argument that a zero-temperature toroidal cloud of gas rotating with speed $v$  would evolve asymptotically into a torus of radius $v t$ after expansion time $t$ implies that central hole radius is a measure of rotational speed (the argument carries over to the quantum domain because the position of the first maximum of $J_{n}(x)^{2}$ varies nearly linearly with $n$).  Fig.\,\ref{fig:Data}(a) shows six successive TOF images taken with the barrier rotating at 2\,Hz.  It is clear that these images show two different hole sizes - corresponding to two different rotational speeds - illustrating directly that while the outcome of the experiment is not always the same the rotational speed of the toroidal BEC is quantized and the system always has integer phase winding, as predicted by Eq.\,(\ref{eq:circulation}).  Analysis shows that images 2 and 3 correspond to winding number $n=1$, while the other four images correspond to $n=2$.  Repeating the experiment for a range of stirring frequencies allows generation of states from $n=0$ up to $n=5$, as seen in Fig.\,\ref{fig:Data}(b).  We note that discrete rotational states in a toroidal BEC have been seen by us \cite{Ryu2010} and by others \cite{Moulder2012, Wright2013}, but the present work is the first to compare experimental results with theoretical predictions.  Further, the quantized rotation seen here results from creating the condensate in a (non-quantized) rotating trap, so the experiment is a BEC analog of the famous Hess-Fairbank rotating bucket experiment \cite{Hess1967}.

We fit the central fringe of each TOF image to a circular function with Gaussian radial profile to estimate the fringe diameter and the peak absorption on that fringe.  Fig.\,\ref{fig:Analysis}(a) is a histogram of the fringe diameter measurements for TOF images exhibiting a central hole.  The clear peaks in the distribution confirm that circulation in the torus is quantized, with the peaks corresponding to successive winding numbers from $n=1$ to $n=5$.  While some of the finite peak width in the histogram is a result of small imperfections in the potential, a scatter plot of peak absorption versus fringe radius [Fig.\,\ref{fig:Analysis}(b)] shows that most of the apparent peak width is associated with a correlation between fitted peak absorption (proportional to atom number for each n) and fitted fringe diameter.  The red lines added to Fig.\,\ref{fig:Analysis}(b) to guide the eye show that an extrapolation of fringe diameter to the limit of zero absorption is in good qualitative agreement with the simple non-interacting Bessel function prediction for the radius of the first fringe, shown as green lines corresponding to the first maximum in $r$ of ${\left| {{J_n}\left( {\frac{{mR}}{{\hbar t}}r} \right)} \right|^2}$.  We believe that the correlation between fringe radius and atom number, which is larger than the interaction effects seen in Fig.\,\ref{fig:GPE-radial}, is an imaging artifact (see text).

For the same data set Fig.\,\ref{fig:Analysis}(c) shows the average winding number as a function of the rotation frequency of the trap.  The procedure of evaporating into the rotating trap followed by a slow lowering of the barrier should find the thermodynamic ground state of the system.  For a trap rotating at angular frequency $\Omega$  about the $z$-axis, that is the state which minimizes the energy associated with the effective Hamiltonian in the rotating frame, $H_{\rm{eff}} = H_{0} - \Omega L_{z}$ where $H_0$ is the hamiltonian for the non-rotating trap and $L_z$ is the $z$ component of the angular momentum.  In our case of a thin toroid, the energy is therefore minimized when the winding number $n$ takes the integer value closest to $\Omega/\Omega_{0}$, giving rise to the equally spaced steps in the theory prediction shown as a red line on Fig.\,\ref{fig:Analysis}(c) (preliminary results from an improved apparatus exhibit these steps and suggest that the method can set the winding number deterministically with fidelity in excess of $90\%$).

\begin{figure}
\includegraphics[width=3.50in]{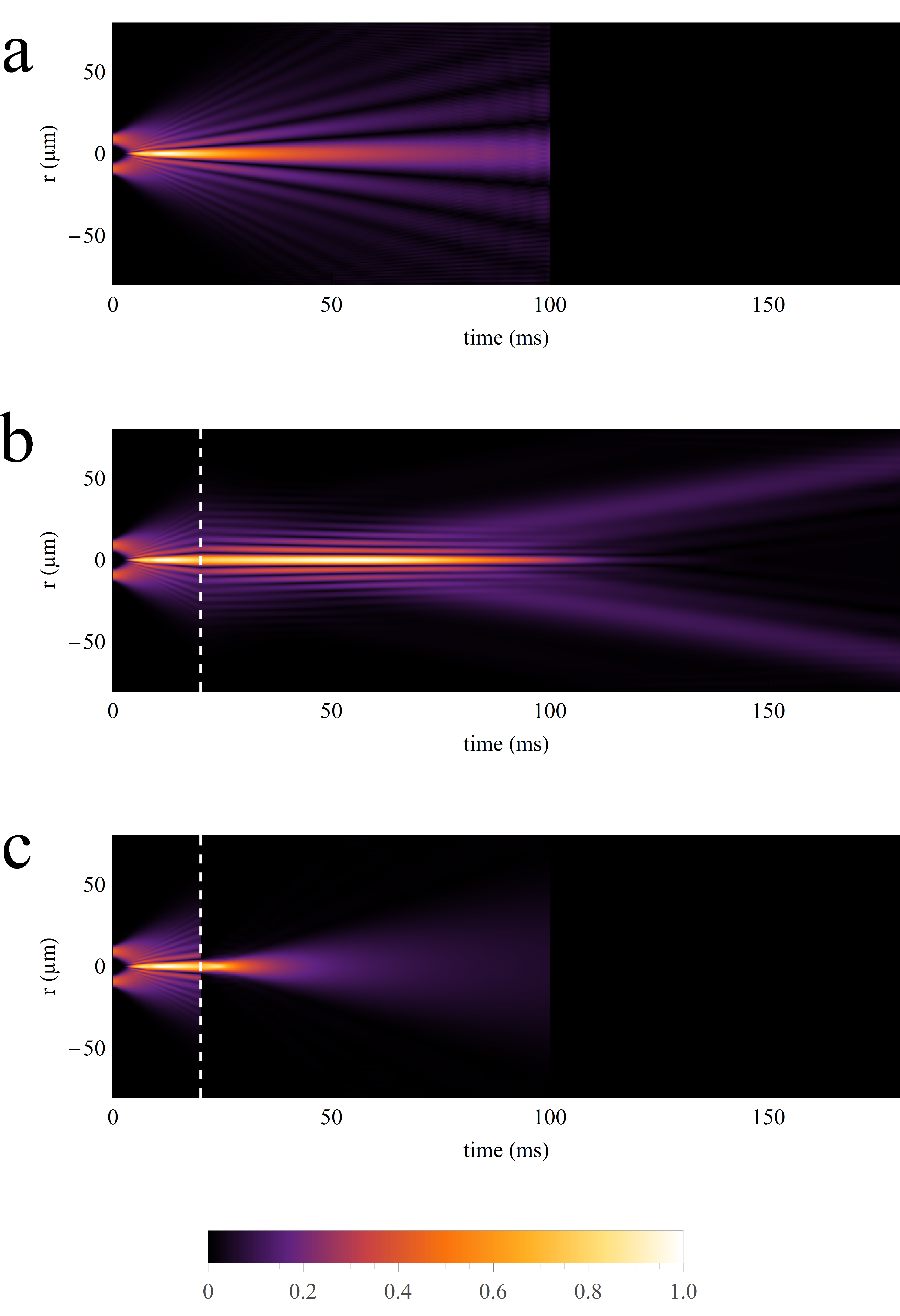}
\caption{\label{fig:GPE-imprinting} Numerical simulation of effect of phase imprinting on the free expansion dynamics of the toroidal BEC, showing time evolution of the normalized column density along the BEC axis of symmetry for winding number $n=0$ obtained from full 3D GPE simulations with parameters appropriate to the experiment.  (a) Free evolution out to time 100\,ms (faint fringing visible after 80\,ms is a numerical artifact).  (b) Quadratic phase (see text) is imprinted on the BEC at time $t =20\,$ms, indicated on the figure by a white dashed line.  (c) Phase imprinted as in (b) and simultaneously masked to allow only the central peak to propagate to time 100\,ms.}
\end{figure}

Fig.\,\ref{fig:GPE-3D} shows that the freely-expanding toroidal condensate eventually forms a very elongated object analogous to a pulsed beam: after 34ms of free evolution the BEC has fallen $6\,$mm under gravity and grown from ~2$\,\mu$m to ~300$\,\mu$m in length.  At the same time, the GPE simulations of the experiment in Fig.\,\ref{fig:GPE-imprinting}(a) and the experimental data in Fig.\,\ref{fig:Data}(b) show that the diameter of the central peak at this point is only $\sim12\,\mu$m, which is still substantially smaller than the original torus.  Because of the correspondingly large aspect ratio and the structure of concentric fringes seen in Fig.\,\ref{fig:GPE-3D}, the six images in Fig.\,\ref{fig:Data}(b) can be regarded as the first experimental demonstration of cold atom matter wave Bessel beams, covering all quantized orbital angular momentum states between $n=0$ and $n=5$.

The GPE simulation in Fig.\,\ref{fig:GPE-imprinting}(a) also shows that the evolution is not completely diffraction-free.  The small divergence is a result of the quadratic phase term in Eq.\,\ref{eq:Fraunhofer}.  Fig.\,\ref{fig:GPE-imprinting}(b) shows the diffraction-free evolution that would be obtained if the condensate had the negative of this phase imprinted on it to remove the curvature (equivalent in optics to collimation by a thin lens).  Because the wavefront in the simulation does not have the infinite Bessel form it propagates diffraction-free for only a finite distance before blowing up, as is well-known from optical implementations \cite{Durnin1987}.  In principle such phase imprinting is easily accomplished experimentally by subjecting the BEC to a short pulse of far-detuned light \cite{Denschlag2000, Burger1999} with a quadratic intensity profile.  However, it was not possible to implement this procedure in our experiment because the poor optical quality of the cell windows prevented delivery of a beam with sufficient intensity smoothness to remove the phase variation of $\sim50\,$radians developed across the condensate after 20\,ms of expansion.  It should be easily demonstrated in future work with an improved system.  Fig.\,\ref{fig:GPE-imprinting}(c) shows the evolution of just the central part of the beam after imprinting combined with masking to suppress the outer fringes.  It propagates almost Heisenberg limited, with calculated beam quality factor \cite{Impens2008} ${M^2} = \sqrt {\left\langle {{x^2}} \right\rangle \left\langle {{p_x}^2} \right\rangle } /(\hbar /2) = 1.1$. It is interesting to note that the presence of the relatively faint outer fringes in the non-imprinted Bessel beam realized in our experiment [Fig.\,\ref{fig:GPE-imprinting}(a)] reduces the divergence of the central beam by a factor $\sim3.5$ compared to the masked beam, increasing the intensity in the central peak (which contains ~$40\%$ of the atoms) by a factor $\sim12$.  So the non-imprinted Bessel beam is, in a sense, better than Heisenberg-limited, and such beams may therefore be preferable to Heisenberg-limited atom lasers \cite{Jeppesen2008} in applications where the additional fringes can be tolerated.

Our experiment can be regarded as a simple form of matter wave holography in which the initial atomic wavefunction is engineered so that the subsequent free evolution (possibly in conjunction with phase imprinting acting as a lens) transforms it into a desired form at some remote location.  Since the Painted Potential can create arbitrary and dynamic potentials there is considerable scope for developing this idea further: the current implementation has about $75 \times 75$ addressable spots, so it should already be possible to create a gray scale hologram with this resolution.  While atom holography has previously been demonstrated with atomic beams passing through a mask serving as a binary transmission hologram \cite{Fujita1996, Morinaga1996}, this form of holography is inefficient because of absorption by the mask, and the hologram cannot be changed once fabricated and installed inside the vacuum system.  Holography with the Painted Potential would consist of forming a suitably shaped condensate by evaporation into a potential landscape corresponding to the desired amplitude distribution in the hologram, or of phase-imprinting a uniform BEC created in a flat box potential.  Painted Potential holography would not waste any atoms through absorption at a mask and it would allow for dynamic and arbitrary holograms.  It could, for example, lead to a new form of microlithography in which surfaces are patterned with atoms in arrangements that can be changed in real time.
 
This work was supported by the U.S. Department of Energy through the LANL/LDRD Program.  We acknowledge inspiring conversations with Eddy Timmermans.

\end{document}